\documentclass[twocolumn,aps,showkeys,amsmath,amssymb]{revtex4}
\usepackage{graphicx}% Include figure files
\usepackage{dcolumn}% Align table columns on decimal point
\usepackage{amsmath}%
\usepackage{amssymb}%

\begin{document}
%\begin{CJK*}{GBK}{song}

%\preprint{APS/000-000}

\title{A theoretical calculation of the cosmological constant based on a mechanical model of vacuum}
% Force line breaks with \\

\author{Xiao-Song Wang}
%\altaffiliation[Also at ]{xxx Department, XYZ University.}
%\email{}
\affiliation{Institute of Mechanical and Power Engineering, Henan Polytechnic University,
Jiaozuo, Henan Province, 454000, China}
\date{December 21, 2019}% It is always \today, today,

\begin{abstract}
Lord Kelvin believed that the electromagnetic aether must also generate gravity. Presently, we have no methods to determine the density of the electromagnetic aether, or we say the $\Omega(1)$ substratum. Thus, we also suppose that vacuum is filled with another kind of continuously distributed substance, which may be called the $\Omega(2)$ substratum. Based on a theorem of V. Fock on the mass tensor of a fluid, the contravariant energy-momentum tensors of the $\Omega(1)$ and $\Omega(2)$ substrata are established. Quasi-static solutions of the gravitational field equations in vacuum are obtained. Based on an assumption, relationships between the contravariant energy-momentum tensors of the $\Omega(1)$ and $\Omega(2)$ substrata and the contravariant metric tensor are obtained. Thus, the cosmological constant is calculated theoretically. The $\Omega(1)$ and $\Omega(2)$ substrata may be a possible candidate of the dark energy. According to the theory of vacuum mechanics, only those energy-momentum tensors of discrete or continuously distributed sinks in the $\Omega(0)$ substratum are permitted to act as the source terms in the generalized Einstein's equations. Thus, the zero-point energy of electromagnetic fields is not qualified for a source term in the generalized Einstein's equations. Some people believed that all kinds of energies should act as source terms in the Einstein's equations. It may be this unwarranted belief that leads to the cosmological constant problem. The mass density of the $\Omega(1)$ and $\Omega(2)$ substrata is equivalent to that of around $3$ protons contained in a box with a volume of $1$ cubic metre.
\end{abstract}


\maketitle

%\tableofcontents

%%%%%%%%%%%%%%%%%%%%%%%%%%%%%%%%%%%%%%%%%%%%%%%%%%%%%%%%%%%%%%%%%%%%%%%%%%
\section{Introduction  \label{sec 100}}
\newtheorem{assumption_my}{\bfseries Assumption}

\newtheorem{definition_my}{\bfseries Definition}

\newtheorem{lemma_my}{\bfseries Lemma}

\newtheorem{proposition_my}{\bfseries Proposition}

\newtheorem{theorem_my}{\bfseries Theorem}

\newtheorem{wcorollary_my}{\bfseries Corollary}

It is known that the cosmological constant problem is one of the puzzles in physics today \cite{ORaifeartaighC2018}. Theoretical interpretation of the small value of the cosmological constant is still open \cite{MarshD2017}.

In 1917, A. Einstein thought that his equations of gravitational fields should be revised to be (\cite{MisnerC1973}, p.\ 410)
\begin{equation}\label{Einstein 100-200}
R_{\mu\nu}-\frac{1}{2}g_{\mu\nu}R + \Lambda g_{\mu\nu} = -\kappa T^{\mathrm{m}}_{\mu\nu},
\end{equation}
where $g_{\mu\nu}$ is the metric tensor of a Riemannian spacetime,
$R_{\mu\nu}$ is the Ricci tensor, $R\equiv g^{\mu\nu}R_{\mu\nu}$ is the scalar curvature, $g^{\mu\nu}$ is the contravariant metric tensor, $\kappa$ is Einstein's gravitational constant, $T^{\mathrm{m}}_{\mu\nu}$ is the energy-momentum tensor of a matter system, $\Lambda$ is the cosmological constant.

It seems that the cosmological constant $\Lambda$ is unnecessary when Hubble discovered the expansion of the universe. Thus, Einstein abandoned the term $\Lambda g_{\mu\nu}$ in Eqs.\ (\ref{Einstein 100-200}) and returned to his original equations (\cite{MisnerC1973}, p.\ 410). Later, the cosmological constant was continuously and intensively studied \cite{ORaifeartaighC2018}.

In 1990-1999 two groups discovered the cosmic vacuum, or dark energy, by studying remote supernova explosions (\cite{ByrdG2012}, p.\ 112). They discovered that some high redshift supernovae appeared fainter and thus more distant than they should be in a gravitationally decelerating universe (\cite{ByrdG2012}, p.\ 113). This discovery gives the first indication that the universe is accelerating. A possible explanation is that the universe may contain some kind of substance which behaves like Einstein's antigravity cosmological constant $\Lambda$ (\cite{ByrdG2012}, p.\ 113).

The term `dark energy' is commonly used to denote a catch-all term for the origin of the observed acceleration of the universe, regardless of whether it arises from a new form of energy or a modification of the general theory of relativity (\cite{ZylaPA2020}, p.\ 490).

The value of the cosmological constant $\Lambda$ is also related to the energy-momentum tensor of vacuum (\cite{MisnerC1973}, p.\ 411). The existence and characters of dark matter and dark energy are still controversy, see for instance  \cite{HamiltonP2015,SpergelD2015,KimJ2017,BosmaA2017,GibneyE2017}.

 The purpose of this paper is to propose a theoretical calculation of the cosmological constant and establish a mechanical model of dark energy based on the theory of vacuum mechanics \cite{WangXS200804,WangXS200810,WangXS2014PhysicsEssays,WangXS201908,WangXS202104}.

\section{A mechanical model of vacuum and a possible origin of the cosmological constant  \label{sec 200}}
Modern experiments, for instance, the Casimir effect \cite{CasimirH1948,SparnaayMJ1957}, have shown that vacuum is not empty. Therefore, new considerations on the old concept of aether may be needed. Vacuum mechanics is a theory attempting to derive some basic physical laws based on a mechanical model of vacuum and elementary particles \cite{WangXS200804,WangXS200810,WangXS2014PhysicsEssays,WangXS201908,WangXS202104}. An introduction of the theory of vacuum mechanics is presented in the Appendix.

An assumption in the theory of vacuum mechanics is that vacuum is filled with a kind of continuously distributed matter which may be called the $\Omega(1)$ substratum, or the electromagnetic aether \cite{WangXS200804}. In 2008, a visco-elastic continuum model of vacuum and a singularity model of electric charges are proposed \cite{WangXS200804}. Then, Maxwell's equations in vacuum are derived by methods of continuum mechanics based on these models \cite{WangXS200804}. Lord Kelvin once said \cite{KelvinL1901}:'{\itshape  That there must be a medium forming a continuous material communication throughout space to the remotest visible body is a fundamental assumption in the undulatory theory of light.}' Lord Kelvin believes that the electromagnetic aether must also generate gravity \cite{KelvinL1901,DupreMJ2012}.

Following Lord Kelvin, we propose an assumption that the particles that constitute the $\Omega(1)$ substratum, are sinks in the $\Omega(0)$ substratum \cite{WangXS201908,WangXS202104}. Thus, according to a previous theory of gravity \cite{WangXS200810}, the $\Omega(1)$ substratum will also generate gravity. Generalized Einstein's equations in a special class of non-inertial reference frames are derived based on the theory of vacuum mechanics \cite{WangXS202104}. If the reference frame is quasi-inertial and the gravitational field is weak, then the generalized Einstein's equations reduce to \cite{WangXS202104}
\begin{equation}\label{Einstein 150-100}
R_{\mu\nu}-\frac{1}{2}g_{\mu\nu}R = \frac{\kappa}{g}\left ( T^{\mathrm{m}}_{\mu\nu}+T^{\Omega(1)}_{\mu\nu}\right ),
\end{equation}
where $T^{\mathrm{m}}_{\mu\nu}$ and $T^{\Omega(1)}_{\mu\nu}$ are the energy-momentum tensors of the matter system and the $\Omega(1)$ substratum respectively,
$g \equiv \mathrm{Det} \ g_{\mu\nu}$.

Comparing Eqs.\ (\ref{Einstein 150-100}) and Eqs.\ (\ref{Einstein 100-200}), it is possible that the term $(\kappa/g)T^{\Omega(1)}_{\mu\nu}$ in Eqs.\ (\ref{Einstein 150-100}) is the origin of the cosmological term $\Lambda g_{\mu\nu}$ in Eqs.\ (\ref{Einstein 100-200}). According to the theory of vacuum mechanics \cite{WangXS200804}, light is the transverse wave of the $\Omega(1)$ substratum. Thus, the velocity of light in vacuum is $c=\sqrt{W/\rho_{1}}$, where $W$ is the shear modulus of the $\Omega(1)$ substratum, $\rho_{1}$ is the mass density of the $\Omega(1)$ substratum \cite{WangXS200804}. Since $c = 2.99792458\times 10^{8}\mathrm{m}\cdot\mathrm{s}^{-1}$ is the largest velocity of transverse elastic waves in elastic media, it is possible that the mass density $\rho_{1}$ of the $\Omega(1)$ substratum may be the smallest of elastic media in the universe. Unfortunately, presently we have no observational data of the density $\rho_{1}$. A bold speculation is that the mass density $\rho_{1}$ may be of the same order of magnitude of the cosmic microwave background (CMB) radiation density $\rho_{\mathrm{cmb}}$ of the universe. However, the observational value of the CMB radiation density is $\rho_{\mathrm{cmb}}=5.38(15)\times 10^{-5} \cdot \rho_{\mathrm{c}}$, where $\rho_{\mathrm{c}} = 8.545(5)\times 10^{-27} \mathrm{kg}\cdot\mathrm{m}^{-3}$ is the critical mass density of the universe (\cite{ZylaPA2020}, p.\ 138). This value of $\rho_{\mathrm{cmb}}$ is too small comparing to the observational value of the mass density $\rho_{\Lambda}=5.83(16)\times 10^{-27} \mathrm{kg}\cdot\mathrm{m}^{-3}$ corresponding to the cosmological constant $\Lambda$ (\cite{ZylaPA2020}, p.\ 138). Therefore, we have to explore other possible methods to explain the origin of the cosmological constant $\Lambda$.

In 2005, C. G. B\"ohmer and T. Harko showed that in the framework of the classical general relativity the presence of a positive cosmological constant implies the existence of a minimal mean density \cite{BohmerCG2005}
\begin{equation}\label{minimal 150-50}
\rho_{\mathrm{min}} = \frac{\Lambda c^{2}}{16\pi G},
\end{equation}
where $c$ is the velocity of light in vacuum, $G$ is Newton's gravitational constant. A bold conjecture is that the mass density $\rho_{1}$ of the $\Omega(1)$ substratum is exactly $\rho_{\mathrm{min}}$, i.e. $\rho_{1}=\Lambda c^{2}/(16\pi G)$. However, the mass density $\rho_{\Lambda}$ corresponding to the cosmological constant $\Lambda$ is \cite{ZeldovichYB1968,SahniV2008}
\begin{equation}\label{density 150-80}
\rho_{\Lambda} = \frac{\Lambda c^{2}}{8\pi G}.
\end{equation}

Comparing Eq.\ (\ref{density 150-80}) and Eq.\ (\ref{minimal 150-50}), $\rho_{\mathrm{min}}$ is exactly $1/2$ of the density $\rho_{\Lambda}$. This minimal mass density $\rho_{\mathrm{min}}$ is not sufficient to explain the mass density $\rho_{\Lambda}$ corresponding to the cosmological constant $\Lambda$.

Therefore, there may exist the following three possibilities. The first possibility is that the mass density $\rho_{1}$ of the $\Omega(1)$ substratum is exactly $\rho_{\Lambda}$, i.e. $\rho_{1}=\Lambda c^{2}/(8\pi G)$. The second possibility is that, except the $\Omega(0)$ and $\Omega(1)$ substratum, there exists a third kind of substratum forming a continuous medium throughout the universe. The mass density of the third kind of substratum will also contribute to $\Lambda$. The third possibility is that there are no other kinds of substrata, except the $\Omega(0)$ and $\Omega(1)$ substratum. Except the contribution of the $\Omega(1)$ substratum to the cosmological constant $\Lambda$, there may exist other unknown reasons which also contribute to $\Lambda$.

Presently, we have no methods to determine the density $\rho_{1}$ of the $\Omega(1)$ substratum. Thus, we cannot exclude or conclude the existence of a third kind of continuously distributed medium in the universe. Therefore, in this manuscript, we focus on the second possibility. Thus, we tentatively introduce the following assumption.
\begin{assumption_my}\label{assumption 150-100}
Vacuum is filled with a kind of continuously distributed substance which may be called the $\Omega(2)$ substratum. The particles that constitute the $\Omega(2)$ substratum may be called the $\Omega(2)$ particles. The $\Omega(2)$ particles are sinks in the $\Omega(0)$ substratum.
\end{assumption_my}

Assumption \ref{assumption 150-100} together with the assumptions in the Appendix provide a mechanical model of vacuum and elementary particles. According to the theory of vacuum mechanics, there exist gravitational interactions between two sinks in the $\Omega(0)$ substratum \cite{WangXS200810}. The $\Omega(1)$ particles, the $\Omega(2)$ particles and elementary particles are sinks in the $\Omega(0)$ substratum \cite{WangXS202104}. Thus, there exist gravitational interactions between the $\Omega(1)$ particles, the $\Omega(2)$ particles and elementary particles. Therefore, the energy-momentum tensor $T^{\Omega(2)}_{\mu\nu}$ of the $\Omega(2)$ substratum will also appear in Eqs.\ (\ref{Einstein 150-100}). Thus, Eqs.\ (\ref{Einstein 150-100}) are revised to be
\begin{equation}\label{Einstein 150-200}
R_{\mu\nu}-\frac{1}{2}g_{\mu\nu}R = \frac{\kappa}{g}\left ( T^{\mathrm{m}}_{\mu\nu}+T^{\Omega(1)}_{\mu\nu}+T^{\Omega(2)}_{\mu\nu}\right ).
\end{equation}

Comparing Eqs.\ (\ref{Einstein 150-200}) and Eqs.\ (\ref{Einstein 100-200}), it is a natural idea that the cosmological term $\Lambda g_{\mu\nu}$ in Eqs.\ (\ref{Einstein 100-200}) may stem from the term $(\kappa/g)\left ( T^{\Omega(1)}_{\mu\nu}+T^{\Omega(2)}_{\mu\nu}\right )$ in Eqs.\ (\ref{Einstein 150-200}). We will investigate this clue in the following sections.

\section{Gravitational field equations in vacuum \label{sec 800}}
The movements of continuously distributed matters
and fields should be studied based on the special theory of relativity \cite{MollerC1972,WeinbergS1972,RindlerW1982}.
Maxwell's equations are valid in the frame of reference that is attached to the $\Omega(1)$ substratum \cite{WangXS200804}.
In order to compare our theory with Newtonian theory of gravitation, which is nonrelativistic,
it is better to single out the speed of light $c$ as a large parameter.
Thus, we introduce the following notation
\begin{equation}\label{notation 800-100}
x^{0} = t, \quad \ x^{1} = x, \quad \  x^{2} = y, \quad \  x^{3} = z,
\end{equation}
where $\{0, x, y, z\}$ is a Cartesian coordinate system
for a three-dimensional Euclidean space that is attached to the $\Omega(1)$ substratum,
$\{0, t\}$ is a one-dimensional time coordinate.
We denote this reference frame as $S_{0}$. If there are no forces acting on a macroscopic body in the $\Omega(1)$ substratum, then the motion of the body is uniform and rectilinear. Thus, the reference frame $S_{0}$ is an inertial frame in the mechanical sense (\cite{FockV1964}, p.\ 15-16; \cite{MollerC1972}, p.\ 2).

We will use Greek indices $\alpha, \beta, \mu, \nu$, etc., for the range $0, 1, 2, 3$
and use Latin indices $i, j, k$, etc., for the range $1, 2, 3$.
According to a theorem of V. Fock, the characteristics of the generalized wave equation, or the d'Alembert equation
\begin{equation}\label{dAlembert 800-200}
\frac{1}{\sqrt{-g}}\frac{\partial }{\partial x^{\alpha}}
\left ( \sqrt{-g}g^{\alpha \beta}\frac{\partial \psi}{\partial x^{\beta}}\right )= 0,
\end{equation}
are (\cite{FockV1964}, p.\ 432)
\begin{equation}\label{characteristics 800-300}
g^{\mu \nu}\frac{\partial \omega}{\partial x^{\mu}}\frac{\partial \omega}{\partial x^{\nu}}= 0,
\end{equation}
where $\psi$ is a field variable, $g_{\mu \nu}$ is a metric tensor,
$g = \mathrm{Det} \ g_{\mu \nu}$, $\omega(x^{0}, x^{1}, x^{2}, x^{3})$ is a characteristic.

Based on the Maxwell's equations, the law of propagation of a characteristic of the electromagnetic wave
in the reference frame $S_{0}$ can be derived and can be written as (\cite{FockV1964}, p.\ 13)
\begin{equation}\label{front 800-400}
\frac{1}{c^{2}}\left ( \frac{\partial \omega}{\partial x^{0}} \right )^{2}-
\left ( \frac{\partial \omega}{\partial x^{1}} \right )^{2}
-\left ( \frac{\partial \omega}{\partial x^{2}} \right )^{2}
-\left ( \frac{\partial \omega}{\partial x^{3}} \right )^{2} = 0,
\end{equation}
where $\omega(x^{0}, x^{1}, x^{2}, x^{3})$ is a characteristic of the electromagnetic wave,
$c$ is the velocity of light in the reference frame $S_{0}$.

Since  equation (\ref{front 800-400}) for a characteristic of the electromagnetic wave in the
$\Omega(1)$ substratum is the same as  equation (\ref{characteristics 800-300}) of the characteristics for the generalized wave equation (\ref{dAlembert 800-200}), the reference frame $S_{0}$ is an inertial frame in the electromagnetic sense (\cite{FockV1964}, p.\ 16). Thus, we define an inertial reference frame as a coordinate system which is static or moving with a constant velocity with respect to the reference frame $S_{0}$ \cite{WangXS201908}.

Comparing Eq.\ (\ref{front 800-400}) and Eq.\ (\ref{characteristics 800-300}), the contravariant metric tensor $g^{\mu\nu}$ must have the following limiting values at infinity (\cite{FockV1964}, p.\ 196)
\begin{eqnarray}
&&(g^{00})_{\infty}= \frac{1}{c^{2}}, \label{metric 800-1510}\\
&&(g^{0i})_{\infty}= 0, \label{metric 800-1520}\\
&&(g^{ij})_{\infty}= -\delta_{ij},\label{metric 800-1530}
\end{eqnarray}
where $\delta_{ij}$ is the Kronecker symbol.

Eqs.\ (\ref{metric 800-1510}-\ref{metric 800-1530}) can be written as
\begin{equation}\label{metric 800-1550}
(g^{\mu\nu})_{\infty} = \left (\begin{array}{rrrr}
  \frac{1}{c^{2}} & 0 & 0 & 0\\
0 & -1 & 0 & 0\\
0 & 0 & -1 & 0\\
0 & 0 & 0 & -1
\end{array}\right ).
\end{equation}

The corresponding limiting values of the metric tensor $g_{\mu\nu}$ will be (\cite{FockV1964}, p.\ 196)
\begin{eqnarray}
&&(g_{00})_{\infty}= c^{2}, \label{metric 800-1610}\\
&&(g_{0i})_{\infty}= 0, \label{metric 800-1620}\\
&&(g_{ij})_{\infty}= -\delta_{ij}.\label{metric 800-1630}
\end{eqnarray}

Eqs.\ (\ref{metric 800-1610}-\ref{metric 800-1630}) can be written as
\begin{equation}\label{metric 800-1700}
(g_{\mu\nu})_{\infty} = \left (\begin{array}{rrrr}
  c^{2} & 0 & 0 & 0\\
0 & -1 & 0 & 0\\
0 & 0 & -1 & 0\\
0 & 0 & 0 & -1
\end{array}\right ).
\end{equation}

Using Eq.\ (\ref{metric 800-1700}), we have $(g_{0})_{\infty} \equiv \mathrm{Det} \ (g_{\mu\nu})_{\infty}=-c^{2}$. Thus, for weak fields, we have $g_{0} \approx (g_{0})_{\infty} = -c^{2}$, where $g_{0}$ is defined by $g_{0} \equiv \mathrm{Det} \ g_{\mu\nu}$ in the coordinate system $S_{0} \equiv \{t, x, y, z\}$.

Since the Einstein's gravitational constant $\kappa$ in Eqs.\ (\ref{Einstein 150-200}) is not a dimensionless constant, $\kappa$ depends on the choice of coordinate system. In order to determine the Einstein's gravitational constant $\kappa$ in the coordinate system $S_{0} \equiv \{t, x, y, z\}$, we study a weak gravitational field of a matter system.  The relationship between the contravariant energy-momentum tensor $T_{\mathrm{m}}^{\mu\nu}$ and the contravariant mass tensor $M_{\mathrm{m}}^{\mu\nu}$ of the matter system is
\begin{equation}\label{relationship 800-1800}
T_{\mathrm{m}}^{\mu\nu}=M_{\mathrm{m}}^{\mu\nu}c^{2}.
\end{equation}

Neglecting the terms $T^{\Omega(1)}_{\mu\nu}$ and $T^{\Omega(2)}_{\mu\nu}$ in Eqs.\ (\ref{Einstein 150-200}) and using $g=g_{0} \approx (g_{0})_{\infty} = -c^{2}$ and Eqs.\ (\ref{relationship 800-1800}), Eqs.\ (\ref{Einstein 150-200}) can be written as
\begin{equation}\label{Einstein 800-1900}
R^{\mu\nu}-\frac{1}{2}g^{\mu\nu}R = -\kappa_{0} M_{\mathrm{m}}^{\mu\nu},
\end{equation}
where $\kappa_{0}$ denotes the Einstein's gravitational constant $\kappa$ in the coordinate system $S_{0} \equiv \{t, x, y, z\}$.

Applying the same method of V. Fock (\cite{FockV1964}, p.\ 198-199), we have
\begin{equation}\label{kappa 800-2000}
\kappa_{0}=\frac{8\pi G}{c^{2}},
\end{equation}
where $G$ is Newton's gravitational constant.

Since there are no atoms in vacuum, we have $T^{\mathrm{m}}_{\mu\nu}=0$. Applying the rules of raising the indices of tensors, i.e. $R_{\alpha\beta}=g_{\alpha\mu}g_{\beta\nu}R^{\mu\nu}$, $g_{\alpha\beta}=g_{\alpha\mu}g_{\beta\nu}g^{\mu\nu}$, $T^{\Omega(1)}_{\alpha\beta}=g_{\alpha\mu}g_{\beta\nu}T_{\Omega(1)}^{\mu\nu}$, $T^{\Omega(2)}_{\alpha\beta}=g_{\alpha\mu}g_{\beta\nu}T_{\Omega(2)}^{\mu\nu}$, the field equations (\ref{Einstein 150-200}) in vacuum reduce to
\begin{equation}\label{Einstein 800-1400}
R^{\mu\nu}-\frac{1}{2}g^{\mu\nu}R = \frac{\kappa_{0}}{g_{0}} \left ( T_{\Omega(1)}^{\mu\nu}+T_{\Omega(2)}^{\mu\nu}\right ).
\end{equation}

\section{Energy-momentum tensor of the $\Omega(1)$ substratum \label{sec 900}}
The Maxwell's equations (\ref{Maxwell 2000-1120}-\ref{Maxwell 2000-1140}) in vacuum can be derived based on a visco-elastic continuum model of the $\Omega(1)$ substratum \cite{WangXS200804}. Let $T_{1}$ be the characteristic time of a macroscopic observer of the $\Omega(1)$ substratum. We may suppose that the observer's time scale $T_{1}$ is very large comparing to the Maxwellian relaxation time $\tau_{1}$ of the $\Omega(1)$ substratum. Therefore, the macroscopic observer concludes that the $\Omega(1)$ substratum behaves like a Newtonian-fluid \cite{WangXS200804}. Thus, the contravariant mass tensor $M_{\Omega(1)\mathrm{S1}}^{\mu\nu}$ of the $\Omega(1)$ substratum in the coordinate system $S_{1} \equiv \{ct, x, y, z\}$ can be written approximatively as (\cite{FockV1964}, p.\ 105)
\begin{eqnarray}
&&M_{\Omega(1)\mathrm{S1}}^{00}=\rho_{1} + \frac{1}{c^{2}}
\left ( \frac{1}{2}\rho_{1} v_{1}^{2} + \rho_{1} \Pi_{1} \right ), \label{mass 900-110}\\
&&M_{\Omega(1)\mathrm{S1}}^{0i}=\frac{\rho_{1}v_{1i}}{c}+\frac{v_{1i}}{c^{3}}\left (\frac{\rho_{1} v_{1}^{2}}{2}+\rho_{1}\Pi_{1} +p_{1}\right ),
\label{mass 900-120}\\
&&M_{\Omega(1)\mathrm{S1}}^{ij}=\frac{1}{c^{2}}\left (\rho_{1} v_{1i}v_{1j}+p_{1}\delta_{ij}\right ),\label{mass 900-130}
\end{eqnarray}
where $\rho_{1}$ is the rest mass density of a volume element of the $\Omega(1)$ substratum
in a laboratory frame relative to which the volume element moving
with velocity $\mathbf{v}_{1}$, $v_{1}=\sqrt{\mathbf{v}_{1}\cdot\mathbf{v}_{1}}$,
$p_{1}$ is the pressure of the $\Omega(1)$ substratum,
\begin{equation}\label{Pi 900-140}
\Pi_{1}=-\frac{p_{1}}{\rho_{1}^{*}}+\int_{0}^{p_{1}}\frac{1}{\rho_{1}^{*}}dp_{1},
\end{equation}
\begin{equation}\label{rho 900-150}
\rho_{1}^{*}=\rho_{1}\sqrt{1-\frac{v_{1}^{2}}{c^{2}}}.
\end{equation}

The relationships between the contravariant mass tensor $M_{\Omega(1)\mathrm{S1}}^{\mu\nu}$ in the coordinate system $S_{1} \equiv \{ct, x, y, z\}$ and the contravariant mass tensor $M_{\Omega(1)}^{\mu\nu}$ in the present coordinate system $S_{0} \equiv \{t, x, y, z\}$ are (\cite{FockV1964}, p.\ 198)
\begin{eqnarray}
&&M_{\Omega(1)}^{00}= \frac{1}{c^{2}}M_{\Omega(1)\mathrm{S1}}^{00}, \label{relationship 900-152}\\
&&M_{\Omega(1)}^{0i}=\frac{1}{c}M_{\Omega(1)\mathrm{S1}}^{0i}, \label{relationship 900-154}\\
&&M_{\Omega(1)}^{ij}=M_{\Omega(1)\mathrm{S1}}^{ij}.\label{relationship 900-156}
\end{eqnarray}

Thus, the contravariant mass tensor $M_{\Omega(1)}^{\mu\nu}$ of the $\Omega(1)$ substratum in the coordinate system $S_{0} \equiv \{t, x, y, z\}$ can be written as
\begin{eqnarray}
&&M_{\Omega(1)}^{00}= \frac{\rho_{1}}{c^{2}} + \frac{1}{c^{4}}\left (\frac{1}{2}\rho_{1} v_{1}^{2} + \rho_{1} \Pi_{1}\right ), \label{fluid 900-210}\\
&&M_{\Omega(1)}^{0i}=\frac{\rho_{1}v_{1i}}{c^{2}}+\frac{v_{1i}}{c^{4}}\left (\frac{\rho_{1} v_{1}^{2}}{2}+\rho_{1}\Pi_{1} +p_{1}\right ),
\label{fluid 900-220}\\
&&M_{\Omega(1)}^{ij}=\frac{1}{c^{2}}\left (\rho_{1} v_{1i}v_{1j}+p_{1}\delta_{ij}\right ).\label{fluid 900-230}
\end{eqnarray}

Then, for the present coordinate system $S_{0} \equiv \{t, x, y, z\}$, the contravariant energy-momentum tensor $T_{\Omega(1)}^{\mu\nu}=M_{\Omega(1)}^{\mu\nu}c^{2}$ can be written as
\begin{eqnarray}
&&T_{\Omega(1)}^{00}= \rho_{1} + \frac{1}{c^{2}}\left (\frac{1}{2}\rho_{1} v_{1}^{2} + \rho_{1} \Pi_{1}\right ), \label{fluid 800-710}\\
&&T_{\Omega(1)}^{0i}=\rho_{1}v_{1i}+\frac{v_{1i}}{c^{2}}\left (\frac{\rho_{1} v_{1}^{2}}{2}+\rho_{1}\Pi_{1} +p_{1}\right ),
\label{fluid 800-720}\\
&&T_{\Omega(1)}^{ij}=\rho_{1} v_{1i}v_{1j}+p_{1}\delta_{ij}.\label{fluid 800-730}
\end{eqnarray}

\begin{assumption_my}\label{assumption 800-800}
Suppose that
\begin{equation}\label{average 800-810}
\langle \rho_{1} v_{1i}v_{1j}+p_{1}\delta_{ij} \rangle = \alpha_{0}\phi_{1}\delta_{ij},
\end{equation}
where $\langle \ \rangle$ denotes the averaged macroscopic value in the sense of statistical physics
from the view point of a macroscopic observer, $\phi_{1}$ is the energy density of a volume element of the $\Omega(1)$ substratum, $\alpha_{0}$ is a parameter to be determined.
\end{assumption_my}

Einstein's mass energy formula for a particle can be written as
\begin{equation}\label{Einstein 800-900}
E = m c^{2},
\end{equation}
where $E$ is the energy of the particle, $m$ is the mass of the particle, $c$ is the velocity of light in vacuum.

Applying Einstein's formula (\ref{Einstein 800-900}) to a volume element of the $\Omega(1)$ substratum, we have
\begin{equation}\label{energy 800-1000}
\phi_{1} = \frac{mc^{2}}{V}= \frac{m_{\mathrm{r}}c^{2}}{V\sqrt{1-\frac{v_{1}^{2}}{c^{2}}}} = \frac{\rho_{1} c^{2}}{\sqrt{1-\frac{v_{1}^{2}}{c^{2}}}},
\end{equation}
where $V$ is the volume of the volume element, $m_{\mathrm{r}}$ is the rest mass of the volume element,
$v_{1}$ is the velocity of the volume element.

Using Eq.\  (\ref{fluid 800-730}), Eq.\  (\ref{average 800-810}) and Eq.\  (\ref{energy 800-1000}), we have
\begin{equation}\label{average 800-1100}
\langle T_{\Omega(1)}^{ij} \rangle = \frac{\alpha_{0}\rho_{1}c^{2}}{\sqrt{1-\frac{v_{1}^{2}}{c^{2}}}}\delta_{ij}, \quad
i, j=1, 2, 3.
\end{equation}

Suppose that $v_{1}^{2} \ll c^{2}$. Thus, $v_{1}^{2}/c^{2} \approx 0$.
Neglecting those terms of order of $1/c^{2}$ in Eq.\  (\ref{fluid 800-710}), Eqs.\ (\ref{fluid 800-720}) and Eqs.\ (\ref{average 800-1100}) and using $\langle T_{\Omega(1)}^{ij} \rangle \approx T_{\Omega(1)}^{ij}$ in Eqs.\ (\ref{average 800-1100}), a macroscopic observer will obtain
\begin{eqnarray}
&&T_{\Omega(1)}^{00} \approx \rho_{1}, \label{fluid 800-1210}\\
&&T_{\Omega(1)}^{0i} \approx \rho_{1}v_{1i}, \label{fluid 800-1220}\\
&&T_{\Omega(1)}^{ij} \approx \alpha_{0}\rho_{1}c^{2}\delta_{ij}. \label{fluid 800-1230}
\end{eqnarray}

Using Eqs.\  (\ref{fluid 800-1210}-\ref{fluid 800-1230}) and Eqs.\ (\ref{metric 800-1610}-\ref{metric 800-1630}), we have
\begin{equation}\label{average 800-1300}
T_{\Omega(1)} = g_{\alpha \beta}T_{\Omega(1)}^{\alpha\beta} \approx (1-3\alpha_{0})\rho_{1}c^{2}.
\end{equation}

\section{Energy-momentum tensor of the $\Omega(2)$ substratum \label{sec 950}}
Let $T_{2}$ be the characteristic time of a macroscopic observer of the $\Omega(2)$ substratum. We may suppose that the observer's time scale $T_{2}$ is very large comparing to the relaxation time $\tau_{2}$ of the $\Omega(2)$ substratum. Therefore, the macroscopic observer concludes that the $\Omega(2)$ substratum behaves like a fluid. Similarly to the case of the $\Omega(1)$ substratum, the contravariant energy-momentum tensor $T_{\Omega(2)}^{\mu\nu}$ of the $\Omega(2)$ substratum in the present coordinate system $S_{0} \equiv \{t, x, y, z\}$ can be written as
\begin{eqnarray}
&&T_{\Omega(2)}^{00}= \rho_{2} + \frac{1}{c^{2}}\left (\frac{1}{2}\rho_{2} v_{2}^{2} + \rho_{2} \Pi_{2}\right ), \label{fluid 950-710}\\
&&T_{\Omega(2)}^{0i}=\rho_{2}v_{2i}+\frac{v_{2i}}{c^{2}}\left (\frac{\rho_{2} v_{2}^{2}}{2}+\rho_{2}\Pi_{2} +p_{2}\right ),
\label{fluid 950-720}\\
&&T_{\Omega(2)}^{ij}=\rho_{2} v_{2i}v_{2j}+p_{2}\delta_{ij},\label{fluid 950-730}
\end{eqnarray}
where $\rho_{2}$ is the rest mass density of a volume element of the $\Omega(2)$ substratum
in a laboratory frame relative to which the volume element moving
with velocity $\mathbf{v}_{2}$, $v_{2}=\sqrt{\mathbf{v}_{2}\cdot\mathbf{v}_{2}}$,
$p_{2}$ is the pressure of the $\Omega(2)$ substratum,
\begin{equation}\label{Pi 950-140}
\Pi_{2}=-\frac{p_{2}}{\rho_{2}^{*}}+\int_{0}^{p_{2}}\frac{1}{\rho_{2}^{*}}dp_{2},
\end{equation}
\begin{equation}\label{rho 950-150}
\rho_{2}^{*}=\rho_{2}\sqrt{1-\frac{v_{2}^{2}}{c^{2}}}.
\end{equation}

We speculate that the macroscopic behaviors of the $\Omega(2)$ substratum are similar to that of the $\Omega(1)$ substratum. Thus, similarly to Assumption \ref{assumption 800-800}, we introduce the following assumption.
\begin{assumption_my}\label{assumption 950-800}
Suppose that
\begin{equation}\label{average 950-810}
\langle \rho_{2} v_{2i}v_{2j}+p_{2}\delta_{ij} \rangle = \alpha_{0}\phi_{2}\delta_{ij},
\end{equation}
where $\phi_{2}$ is the energy density of a volume element of the $\Omega(2)$ substratum.
\end{assumption_my}

Similarly to the case of the $\Omega(1)$ substratum, we have
\begin{equation}\label{average 950-1100}
\langle T_{\Omega(2)}^{ij} \rangle = \frac{\alpha_{0}\rho_{2}c^{2}}{\sqrt{1-\frac{v_{2}^{2}}{c^{2}}}}\delta_{ij}, \quad
i, j=1, 2, 3.
\end{equation}

Suppose that $v_{2}^{2} \ll c^{2}$. Thus, $v_{2}^{2}/c^{2} \approx 0$. Neglecting those terms of order of $1/c^{2}$ in Eq.\  (\ref{fluid 950-710}), Eqs.\ (\ref{fluid 950-720}) and Eqs.\ (\ref{average 950-1100}) and using $\langle T_{\Omega(2)}^{ij} \rangle \approx T_{\Omega(2)}^{ij}$ in Eqs.\ (\ref{average 950-1100}), a macroscopic observer will obtain
\begin{eqnarray}
&&T_{\Omega(2)}^{00} \approx \rho_{2}, \label{fluid 950-1210}\\
&&T_{\Omega(2)}^{0i} \approx \rho_{2}v_{2i}, \label{fluid 950-1220}\\
&&T_{\Omega(2)}^{ij} \approx \alpha_{0}\rho_{2}c^{2}\delta_{ij}. \label{fluid 950-1230}
\end{eqnarray}

Using Eq.\  (\ref{fluid 950-1210}-\ref{fluid 950-1230}) and Eqs.\ (\ref{metric 800-1610}-\ref{metric 800-1630}), we have
\begin{equation}\label{average 950-1300}
T_{\Omega(2)} = g_{\alpha \beta}T_{\Omega(2)}^{\alpha\beta} \approx (1-3\alpha_{0})\rho_{2}c^{2}.
\end{equation}

\section{Quasi-static solutions of the gravitational field equations in vacuum \label{sec 1000}}
Since the field is weak, we have $g_{0} \approx -c^{2}$.  Thus, the field equations (\ref{Einstein 800-1400}) can be written as (\cite{FockV1964}, p.\ 198)
\begin{equation}\label{Einstein 800-1500}
R^{\mu\nu}=-\frac{\kappa_{0}}{c^{2}}\left[T_{\Omega(1)}^{\mu\nu}+T_{\Omega(2)}^{\mu\nu}-\frac{1}{2}g^{\mu\nu}\left (T_{\Omega(1)}+T_{\Omega(2)}\right )\right ].
\end{equation}

Using Eqs.\ (\ref{metric 800-1510}-\ref{metric 800-1530}), Eqs.(\ref{fluid 800-1210}-\ref{average 800-1300}) and Eqs.(\ref{fluid 950-1210}-\ref{average 950-1300}), we have
\begin{eqnarray}
T_{\Omega(1)}^{00}+T_{\Omega(2)}^{00}&-&\frac{1}{2}g^{00}\left (T_{\Omega(1)}+T_{\Omega(2)}\right )\nonumber \\
 && \approx \frac{(1+3\alpha_{0})(\rho_{1}+\rho_{2})}{2}, \label{elastic 800-2110}\\
T_{\Omega(1)}^{0i}+T_{\Omega(2)}^{0i}&-&\frac{1}{2}g^{0i}\left (T_{\Omega(1)}+T_{\Omega(2)}\right )\nonumber \\
 && \approx \rho_{1}v_{1i}+\rho_{2}v_{2i}, \label{elastic 800-2120}\\
T_{\Omega(1)}^{ij}+T_{\Omega(2)}^{ij}&-&\frac{1}{2}g^{ij}\left (T_{\Omega(1)}+T_{\Omega(2)}\right )\nonumber \\
 && \approx \frac{(1-\alpha_{0})(\rho_{1}+\rho_{2})c^{2}}{2}\delta_{ij}. \label{elastic 800-2130}
\end{eqnarray}

An inertial reference frame is harmonic (\cite{FockV1964}, p.\ 370). For harmonic coordinates, we have approximately (\cite{FockV1964}, p.\ 198)
\begin{equation}\label{Ricci 1000-100}
R^{\mu\nu}=-\frac{1}{2c^{2}}\frac{\partial^{2}g^{\mu\nu}}{\partial t^{2}} + \frac{1}{2}\nabla^{2}g^{\mu\nu},
\end{equation}
where $\nabla^{2} = \partial^{2} /\partial x^{2} + \partial^{2} /\partial y^{2} +\partial^{2} /\partial z^{2}$ is the Laplace operator, i.e. \begin{equation}\label{dAlembert 1000-200}
\nabla^{2}g^{\mu\nu}
= \frac{\partial^{2} g^{\mu\nu}}{\partial (x^{1})^{2}}
+\frac{\partial^{2} g^{\mu\nu}}{\partial (x^{2})^{2}}
+\frac{\partial^{2} g^{\mu\nu}}{\partial (x^{3})^{2}}.
\end{equation}

Since we are interested in the quasi-static behaviors of $g^{\mu\nu}$, we suppose that (\cite{FockV1964}, p.\ 198)
\begin{equation}\label{assumption 1000-300}
\frac{\partial^{2}g^{\mu\nu}}{c^{2}\partial t^{2}} \approx 0.
\end{equation}

Putting Eqs.(\ref{elastic 800-2110}-\ref{Ricci 1000-100})
into Eqs.(\ref{Einstein 800-1500}) and using Eqs.\ (\ref{assumption 1000-300}), we obtain
\begin{eqnarray}
&&\nabla^{2}g^{00}
\approx -\frac{\kappa_{0}}{c^{2}}(1+3\alpha_{0})(\rho_{1}+\rho_{2}), \label{field 1000-410}\\
&&\nabla^{2}g^{0i} \approx -\frac{2\kappa_{0}}{c^{2}}(\rho_{1}v_{1i}+\rho_{2}v_{2i}), \label{field 1000-420}\\
&&\nabla^{2}g^{ij}
\approx  -\kappa_{0}(1-\alpha_{0})(\rho_{1}+\rho_{2})\delta_{ij}. \label{field 1000-430}
\end{eqnarray}

Newton's gravitational potential $U_{0}$ satisfies the following equation (\cite{FockV1964}, p.\ 199)
\begin{equation}\label{potential 800-2400}
\nabla^{2}U_{0} = 4\pi G(\rho_{1}+\rho_{2}).
\end{equation}

Following V. Fock (\cite{FockV1964}, p.\ 199),
we introduce the following gravitational vector potentials $U_{i}$ which satisfy
\begin{equation}\label{potential 800-2500}
\nabla^{2}U_{i} = 4\pi G(\rho_{1}v_{1i}+\rho_{2}v_{2i}), \quad i=1, 2, 3.
\end{equation}

\begin{proposition_my}\label{metric 800-2600}
Suppose that the following conditions are valid: (1) $U_{0} \approx 0$ and $U_{i} \approx 0$ at infinity;
(2) in the spherical coordinates $(r,\theta,\psi)$, $U_{0}$ and $g^{\mu\nu}$ are axially symmetric, i.e.
they do not vary with the azimuthal angle $\psi$;
(3) the contravariant metric tensor $g^{\mu\nu}$ depends on $r$
and $\theta$ as the Newton potential $U_{0}$; (4) the law of propagation of an electromagnetic wave front, i.e.
Eq.\  (\ref{front 800-400}), is valid in the reference frame $S_{0}$ at infinity.  Then,
the solutions of Eqs.\  (\ref{field 1000-410}-\ref{field 1000-430}) are
\begin{eqnarray}
&&g^{00} \approx \frac{1}{c^{2}}
-\frac{2(1+3\alpha_{0})U_{0}}{c^{4}},\label{metric 800-2610}\\
&&g^{0i} \approx -\frac{4U_{i}}{c^{4}}, \label{metric 800-2620}\\
&&g^{ij} \approx \left [-1-\frac{2(1-\alpha_{0})U_{0}}{c^{2}}\right ]\delta_{ij}.
\label{metric 800-2630}
\end{eqnarray}
\end{proposition_my}
{\bfseries{Proof.}} From Eq.\  (\ref{potential 800-2400}), we have
\begin{equation}\label{density 800-2610}
\rho_{1} +\rho_{2} = \frac{1}{4\pi G}\nabla^{2}U_{0}.
\end{equation}

Putting Eq.\  (\ref{density 800-2610}) into Eq.\  (\ref{field 1000-410}), we have
\begin{equation}\label{g00 800-2610}
\nabla^{2}(g^{00}+\beta_{0}U_{0})\approx 0,
\end{equation}
where
\begin{equation}\label{parameter 1000-500}
\beta_{0} = \frac{(1+3\alpha_{0})\kappa_{0}}{4\pi Gc^{2}}.
\end{equation}

Using Eq.\  (\ref{kappa 800-2000}), Eq.\  (\ref{parameter 1000-500}) can be written as
\begin{equation}\label{parameter 1000-700}
\beta_{0} = \frac{2(1+3\alpha_{0})}{c^{4}}.
\end{equation}

Eq.\  (\ref{g00 800-2610}) is a Laplace's equation. Since $U_{0}$ and $g^{00}$ are axially symmetric, a general form of solution of Eq.\  (\ref{g00 800-2610}) can be
obtained by using the spherical coordinates $(r,\theta,\varphi)$ (\cite{Currie2003}, p.\ 132)
\begin{equation}\label{solution 800-2620}
g^{00}(r,\theta)+\beta_{0}U_{0}(r,\theta)\approx \sum_{n}\left(
A_{n}r^{n}+\frac{B_{n}}{r^{n+1}}\right)P_{n}(\cos \theta),
\end{equation}
where $A_{n}$ and $B_{n}$ are arbitrary constants, $n$ are integers,
 $P_{n}(x)$ are Legendre's functions of the first kind which are defined by
\begin{equation}\label{Legendre 800-2630}
P_{n}(x)=\frac{1}{2^{n}n!}\frac{\mathrm{d}^{n}}{\mathrm{d}x^{n}}\,(x^{2}-1)^{n}.
\end{equation}

We have assumed that the contravariant metric tensor $g^{\mu\nu}$ depends on $r$ and $\theta$ as the Newton potential $U_{0}$. Thus, we set $n=0$ and $B_{n}=0$ in Eq.\  (\ref{solution 800-2620}) and obtain
\begin{equation}\label{solution 800-2640}
g^{00}+\beta_{0}U_{0}\approx A_{0},
\end{equation}
where $A_{0}$ is a constant to be determined.

Using Eq.\  (\ref{potential 800-2500}) and Eq.\  (\ref{kappa 800-2000}), Eqs.\  (\ref{field 1000-420}) can be written as
\begin{equation}\label{field 1000-2700}
\nabla^{2}\left (g^{0i}+\frac{4}{c^{4}}U_{i}\right )\approx 0.
\end{equation}

Using Eq.\  (\ref{potential 800-2400}) and Eq.\  (\ref{kappa 800-2000}), Eqs.\  (\ref{field 1000-430}) can be written as
\begin{equation}\label{field 1000-2800}
\nabla^{2}\left (g^{ij}+\frac{2(1-\alpha_{0})}{c^{2}}U_{0}\delta_{ij}\right )\approx 0.
\end{equation}

Applying similar method in solving Eq.\  (\ref{g00 800-2610}) to Eqs.\  (\ref{field 1000-2700}) and Eqs.\  (\ref{field 1000-2800}), we have
\begin{eqnarray}
&&g^{0i}+\beta_{1}U_{i}\approx A_{1},\label{metric 1000-710}\\
&&g^{ij}+\beta_{2}U_{0}\delta_{ij}\approx A_{2}\delta_{ij}, \label{metric 1000-720}
\end{eqnarray}
where $A_{1}$ and $A_{2}$ are constants to be determined,
\begin{equation}\label{parameter 1000-750}
\beta_{1} = \frac{4}{c^{4}}, \quad \beta_{2} = \frac{2(1-\alpha_{0})}{c^{2}}.
\end{equation}

Noticing $U_{0} \approx U_{i} \approx 0$ at infinity and using Eqs.\  (\ref{metric 800-1510}-\ref{metric 800-1530}) in Eq.\ (\ref{solution 800-2640}) and Eqs.\  (\ref{metric 1000-710}-\ref{metric 1000-720}), we have
\begin{equation}\label{solution 800-2680}
A_{0} = \frac{1}{c^{2}}, \quad \ A_{1} = 0, \quad \ A_{2} = -1.
\end{equation}

Thus, the solutions of Eqs.\ (\ref{field 1000-410}-\ref{field 1000-430}) are Eq.\ (\ref{solution 800-2640}), Eqs.\ (\ref{metric 1000-710}) and Eqs.\ (\ref{metric 1000-720}), which can be written as Eqs.\ (\ref{metric 800-2610}-\ref{metric 800-2630}). $\Box$

The covariant metric tensor $g_{\mu\nu}$ can be derived from the contravariant metric tensor $g^{\mu\nu}$ and the results are  (\cite{FockV1964}, p.\ 201)
\begin{eqnarray}
&&g_{00} \approx c^{2}+2(1+3\alpha_{0})U_{0},\label{metric 1000-910}\\
&&g_{0i} \approx -\frac{4U_{i}}{c^{4}}, \label{metric 1000-920}\\
&&g_{ij} \approx \left [-1+\frac{2(1-\alpha_{0})U_{0}}{c^{2}}\right ]\delta_{ij}.\label{metric 1000-930}
\end{eqnarray}

\section{A theoretical derivation of the cosmological constant \label{sec 1100}}
We speculate that the $\Omega(1)$ and $\Omega(2)$ substrata may be responsible for the so-called dark energy. Following this idea, we set out to explore the possible value of the parameter $\alpha_{0}$ in Eqs.\ (\ref{average 800-810}) and Eqs.\ (\ref{average 950-810}).

We have supposed that the velocity $v_{1}$ and $v_{2}$ of the volume element of the $\Omega(1)$ substratum, or the $\Omega(2)$ substratum, are small, i.e. $v_{1}^{2} \ll c^{2}$ and $v_{2}^{2} \ll c^{2}$. Thus, the terms $\rho_{1} v_{1i}$ and $\rho_{2} v_{2i}$ in Eqs.\ (\ref{fluid 800-1220}) and Eqs.\ (\ref{fluid 950-1220}) may be neglected. Then, according to Eqs.\ (\ref{fluid 800-1210}-\ref{fluid 800-1230}) and Eqs.\ (\ref{fluid 950-1210}-\ref{fluid 950-1230}), we have
\begin{equation}\label{tensor 1100-100}
T_{\Omega(1)}^{\mu\nu}+T_{\Omega(2)}^{\mu\nu} \approx (\rho_{1}+\rho_{2})c^{2}\left (\begin{array}{rrrr}
  \frac{1}{c^{2}} & 0 & 0 & 0\\
0 & \alpha_{0} & 0 & 0\\
0 & 0 & \alpha_{0} & 0\\
0 & 0 & 0 & \alpha_{0}
\end{array}\right ).
\end{equation}

Since $v_{1}$, $v_{2}$, $\rho_{1}$ and $\rho_{2}$ are small enough, Newton's gravitational potential $U_{0}$ in Eq.\  (\ref{potential 800-2400}) and the gravitational vector potentials $U_{i}$ in Eqs.\ (\ref{potential 800-2500}) are small relative to $c^{2}$. Thus, the terms involving $U_{0}$ or $U_{i}$ in Eqs.\  (\ref{metric 800-2610}-\ref{metric 800-2630}) may be neglected. Then, using Eqs.\  (\ref{metric 800-2610}-\ref{metric 800-2630}), the contravariant metric tensor $g^{\mu\nu}$ of the $\Omega(1)$ substratum can be written as
\begin{equation}\label{metric 1100-200}
g^{\mu\nu} \approx \left (\begin{array}{rrrr}
  \frac{1}{c^{2}} & 0 & 0 & 0\\
0 & -1 & 0 & 0\\
0 & 0 & -1 & 0\\
0 & 0 & 0 & -1
\end{array}\right ).
\end{equation}

In Section 2, we speculate that the cosmological term $\Lambda g_{\mu\nu}$ in Eqs.\ (\ref{Einstein 100-200}) may stem from the term $(\kappa/g)\left ( T^{\Omega(1)}_{\mu\nu}+T^{\Omega(2)}_{\mu\nu}\right )$ in Eqs.\ (\ref{Einstein 150-200}). Now comparing Eq.\  (\ref{tensor 1100-100}) and Eq.\  (\ref{metric 1100-200}), we conjecture that $g^{\mu\nu}$ may depend on $T_{\Omega(1)}^{\mu\nu}+T_{\Omega(2)}^{\mu\nu}$ approximately linearly.

If the general theory of relativity is the correct description of gravity on cosmological scales, then the observed acceleration of the universe requires a new energy component beyond visible matter (\cite{ZylaPA2020}, p.\ 490). The cosmic acceleration could arise from a general form of dark energy that has negative pressure, typically specified by the following equation of state (\cite{ZylaPA2020}, p.\ 490)
\begin{equation}\label{state 1100-320}
p_{\mathrm{vac}}=w\phi_{\mathrm{vac}},
\end{equation}
where $p_{\mathrm{vac}}$ is the pressure of vacuum, $\phi_{\mathrm{vac}}$ is the energy density of vacuum, $w$ is the equation of state parameter.

The observed value of the dark energy equation of state parameter is $w=-1.02831$ (\cite{ZylaPA2020}, p.\ 139).
In 1968, Y. B. Zeldovich calculated the energy density $\phi_{\mathrm{vac}}$ and pressure $p_{\mathrm{vac}}$ of vacuum based on the field theory with relativistically invariant regularisation procedure. He obtained an equation of state of vacuum \cite{ZeldovichYB1968,SahniV2008}
\begin{equation}\label{state 1100-350}
p_{\mathrm{vac}}=-\phi_{\mathrm{vac}},
\end{equation}
where the vacuum pressure $p_{\mathrm{vac}}$ and energy density $\phi_{\mathrm{vac}}$ are finite.

Eq.\  (\ref{state 1100-350}) predicts a value of the dark energy equation of state parameter $w=-1$. This value is consistent with observed value $w=-1.02831$ (\cite{ZylaPA2020}, p.\ 139).

From Eq.\ (\ref{average 800-810}) and Eq.\ (\ref{average 950-810}), we have
\begin{equation}\label{state 1100-360}
\langle \rho_{1}v_{1i}v_{1j}+\rho_{2}v_{2i}v_{2j}+(p_{1}+p_{2})\delta_{ij} \rangle=\alpha_{0}(\phi_{1}+\phi_{2})\delta_{ij}.
\end{equation}

The physical meaning of $\langle \rho_{1}v_{1i}v_{1j}+p_{1}\delta_{ij} \rangle$ is the macroscopic pressure in the $\Omega(1)$ substrata. $\langle \rho_{2}v_{2i}v_{2j}+p_{2}\delta_{ij} \rangle$ is the macroscopic pressure in the $\Omega(2)$ substrata. Thus, the physical meaning of $\langle \rho_{1}v_{1i}v_{1j}+\rho_{2}v_{2i}v_{2j}+(p_{1}+p_{2})\delta_{ij} \rangle$ is the macroscopic pressure tensor $p_{\mathrm{vac}}\delta_{ij}$ in the $\Omega(1)$ and $\Omega(2)$ substrata. Thus, Eq.\  (\ref{state 1100-360}) can be written as
\begin{equation}\label{state 1100-380}
p_{\mathrm{vac}}=\alpha_{0}(\phi_{1}+\phi_{2}).
\end{equation}

Comparing Eq.\ (\ref{state 1100-380}) with Eq.\  (\ref{state 1100-350}) and Eq.\ (\ref{state 1100-320}), we conjecture that $\alpha_{0} = -1$. Thus, inspired by Eq.\  (\ref{state 1100-350}) and Eq.\ (\ref{state 1100-320}), we introduce the following assumption.
\begin{assumption_my}\label{assumption 1100-400}
$\alpha_{0} \approx -1$.
\end{assumption_my}

Applying Assumption \ref{assumption 1100-400},  Eqs.\  (\ref{tensor 1100-100}) can be written as
\begin{equation}\label{tensor 1100-500}
T_{\Omega(1)}^{\mu\nu}+T_{\Omega(2)}^{\mu\nu} \approx (\rho_{1}+\rho_{2})c^{2}\left (\begin{array}{rrrr}
  \frac{1}{c^{2}} & 0 & 0 & 0\\
0 & -1 & 0 & 0\\
0 & 0 & -1 & 0\\
0 & 0 & 0 & -1
\end{array}\right ).
\end{equation}

Putting Eqs.\  (\ref{metric 1100-200}) into Eqs.\  (\ref{tensor 1100-500}), we obtain the following relationships
\begin{equation}\label{relationship 1100-400}
T^{\Omega(1)}_{\mu\nu}+T^{\Omega(2)}_{\mu\nu} \approx (\rho_{1}+\rho_{2})c^{2}g_{\mu\nu}.
\end{equation}

Now we study the gravitational field of a system of matter, the $\Omega(1)$ substratum and the $\Omega(2)$ substratum. We suppose that the gravitational field is weak, i.e. Eq.\  (\ref{metric 1100-200}) is valid. Thus, we have $g_{0} \approx -c^{2}$ and Eqs.\ (\ref{relationship 1100-400}) are valid. Putting Eqs.\ (\ref{relationship 1100-400}) into Eqs.\ (\ref{Einstein 150-200}) and using $g=g_{0} \approx -c^{2}$, we have
\begin{equation}\label{Einstein 1100-500}
R_{\mu\nu}-\frac{1}{2}g_{\mu\nu}R \approx -\frac{\kappa_{0}}{c^{2}} T^{\mathrm{m}}_{\mu\nu}-\kappa_{0} (\rho_{1}+\rho_{2})g_{\mu\nu}.
\end{equation}

Comparing Eqs.\ (\ref{Einstein 1100-500}) and Eqs.\ (\ref{Einstein 100-200}), we introduce the following notation
\begin{eqnarray}
\kappa &=& \frac{\kappa_{0}}{c^{2}}, \label{relationship 1100-610}\\
\Lambda &=& \kappa_{0}(\rho_{1}+\rho_{2}). \label{relationship 1100-620}
\end{eqnarray}

Using Eq.\  (\ref{kappa 800-2000}), Eq.\  (\ref{relationship 1100-620}) can be written as
\begin{equation}\label{relationship 1100-700}
\Lambda = \frac{8\pi G(\rho_{1}+\rho_{2})}{c^{2}}.
\end{equation}

Eq.\  (\ref{relationship 1100-700}) is the theoretical calculation of the cosmological constant based on the theory of vacuum mechanics. Using Eq.\  (\ref{relationship 1100-700}), it is possible for us to calculate the value of $\Lambda$ based on observations. The critical mass density $\rho_{\mathrm{c}}$ of the universe is $\rho_{\mathrm{c}} = 3H_{0}/(8\pi G) = 8.545(5)\times 10^{-27} \mathrm{kg}\cdot\mathrm{m}^{-3}$, where $H_{0}$ is the present-day Hubble expansion rate (\cite{ZylaPA2020}, p.\ 138). The dark energy density parameter is $\Omega_{\Lambda}=0.685(7)$ (\cite{ZylaPA2020}, p.\ 138). Thus, the mass density $\rho_{\Lambda}$ corresponding to the dark energy is $\rho_{\Lambda} = \Omega_{\Lambda}\rho_{\mathrm{c}} = 5.859(7)\times 10^{-27} \mathrm{kg}\cdot\mathrm{m}^{-3}$. According to the mechanical model of vacuum in Section 2, vacuum is filled with two kinds of continuously distributed sinks of the $\Omega(0)$ substrata, i.e. the $\Omega(1)$ and $\Omega(2)$ substrata. Thus, we have
\begin{equation}\label{density 1100-800}
\rho_{\mathrm{vac}} = \rho_{1}+\rho_{2}.
\end{equation}

Suppose that the mass density $\rho_{\mathrm{vac}}$ of vacuum is approximately the mass density $\rho_{\Lambda}$ corresponding to the dark energy, i.e.
\begin{equation}\label{density 1100-850}
\rho_{\mathrm{vac}} \approx \rho_{\Lambda}.
\end{equation}

Comparing Eq.\ (\ref{density 1100-850}) and Eq.\ (\ref{density 1100-800}), we have
\begin{equation}\label{density 1100-900}
\rho_{1}+\rho_{2} \approx \rho_{\Lambda} = 5.859(7)\times 10^{-27} \mathrm{kg}\cdot\mathrm{m}^{-3}.
\end{equation}

We have the following data (\cite{ZylaPA2020}, p.\ 137)
\begin{eqnarray}
c &=& 2.99792458\times 10^{8}\mathrm{m}\cdot\mathrm{s}^{-1}, \label{c 1100-1000}\\
G &=& 6.67430(15)\times 10^{-11}\mathrm{m}^{-2}\mathrm{kg}^{-1}\mathrm{s}^{-2}. \label{G 1100-1100}
\end{eqnarray}

Putting Eq.\  (\ref{density 1100-900}-\ref{G 1100-1100}) into Eq.\  (\ref{relationship 1100-700}), we have the following theoretical value of the cosmological constant $\Lambda_{\mathrm{the}}$
\begin{equation}\label{Lambda 1100-1200}
\Lambda_{\mathrm{the}} = 1.093(65)\times 10^{-52}\mathrm{m}^{-2}.
\end{equation}

The theoretical value of the cosmological constant $\Lambda_{\mathrm{the}}$ in Eq.\  (\ref{Lambda 1100-1200}) is consistent with the observational value of the cosmological constant $\Lambda_{\mathrm{obs}}=1.088(30)\times 10^{-52}\mathrm{m}^{-2}$ (\cite{ZylaPA2020}, p.\ 138).

\section{A theoretical calculation of the mass density of the $\Omega(1)$ and $\Omega(2)$ substrata \label{sec 1150}}
The mass density of the electromagnetic aether is an open problem since eighteenth century \cite{WhittakerE1910,WhittakerE1953,SchaffnerKF1972}.
From Eq.\  (\ref{relationship 1100-700}), we have
\begin{equation}\label{density 1150-100}
\rho_{1}+\rho_{2} = \frac{\Lambda c^{2}}{8\pi G}.
\end{equation}

Noticing that $v_{1}^{2} \ll c^{2}$ and $v_{2}^{2} \ll c^{2}$ and using Eq.\ (\ref{energy 800-1000}) and Eq.\ (\ref{density 1150-100}), the energy density $\phi_{1}+\phi_{2}$ of the $\Omega(1)$ and $\Omega(2)$ substrata is
\begin{equation}\label{density 1150-150}
\phi_{1}+\phi_{2} = (\rho_{1}+\rho_{2})c^{2} = \frac{\Lambda c^{4}}{8\pi G}.
\end{equation}

We have an observational data of the cosmological constant $\Lambda$ (\cite{ZylaPA2020}, p.\ 138)
\begin{equation}\label{data 1150-210}
\Lambda = 1.088(30)\times 10^{-52}\mathrm{m}^{-2}.
\end{equation}

Putting Eq.\  (\ref{c 1100-1000}), Eq.\  (\ref{G 1100-1100}) and Eq.\  (\ref{data 1150-210}) into Eq.\  (\ref{density 1150-100}), we have
\begin{equation}\label{data 1150-310}
\rho_{1}+\rho_{2} = 5.831(02)\times 10^{-27} \mathrm{kg}\cdot\mathrm{m}^{-3}.
\end{equation}

The mass $m_{\mathrm{p}}$ of a proton is (\cite{ZylaPA2020}, p.\ 137)
\begin{equation}\label{data 1150-400}
m_{\mathrm{p}} = 1.67262192369(51)\times 10^{-27} \mathrm{kg}.
\end{equation}

Comparing Eq.\  (\ref{data 1150-400}) and Eq.\  (\ref{data 1150-310}), the mass density $\rho_{1}+\rho_{2}$ of the $\Omega(1)$ and $\Omega(2)$ substrata is equivalent to that of around $3$ protons contained in a box with a volume of $1.0 \mathrm{m}^{3}$. Unfortunately, we have no methods to determine the density $\rho_{1}$ or $\rho_{2}$ individually.

\section{A possible candidate of the dark energy \label{sec 1200}}
In Section 2, a mechanical model of vacuum and elementary particles is established. In this model, the $\Omega(1)$ particles, the $\Omega(2)$ particles and elementary particles are sinks in the $\Omega(0)$ substratum. Thus, according to the theory of vacuum mechanics \cite{WangXS200810}, there exist gravitational interactions between the $\Omega(1)$ or $\Omega(2)$ particles and elementary particles. Therefore, the gravitational fields of the $\Omega(1)$ or $\Omega(2)$ substrata will influence the motions of elementary particles. Thus, the energy-momentum tensors $T^{\Omega(1)}_{\mu\nu}$ and $T^{\Omega(2)}_{\mu\nu}$ of the $\Omega(1)$ and $\Omega(2)$ substrata should also be included in the gravitational field equations (\ref{Einstein 150-100}).

The dark energy is believed to be responsible for the observed acceleration of the universe (\cite{ZylaPA2020}, p.\ 490).
Since the $\Omega(1)$ or $\Omega(2)$ substratum is continuously distributed in vacuum, the gravitational fields of the $\Omega(1)$ or $\Omega(2)$ substrata may behave like the so-called dark energy (\cite{ZylaPA2020}, p.\ 490).
Indeed, according to Eqs.\  (\ref{relationship 1100-400}), Eqs.\  (\ref{Einstein 1100-500}) and Eq.\  (\ref{relationship 1100-620}), we notice that the origin of the cosmological term $\Lambda g_{\mu\nu}$ in Eqs.\ (\ref{Einstein 100-200}) may be the energy-momentum tensors $T^{\Omega(1)}_{\mu\nu}+T^{\Omega(2)}_{\mu\nu}$ of the $\Omega(1)$ and $\Omega(2)$ substrata. Therefore, we speculate that the $\Omega(1)$ and $\Omega(2)$ substrata may be a possible candidate of the so-called concept of dark energy (\cite{ZylaPA2020}, p.\ 490).

Applying Assumption \ref{assumption 1100-400},  Eq.\  (\ref{state 1100-380}) can be written as
\begin{equation}\label{pressure 1200-100}
p_{\mathrm{vac}}=-(\phi_{1}+\phi_{2}).
\end{equation}

Eq.\ (\ref{pressure 1200-100}) shows that the macroscopic pressure $p_{\mathrm{vac}}$ in the $\Omega(1)$ and $\Omega(2)$ substrata is negative. From Eq.\ (\ref{state 1100-320}) and Eq.\ (\ref{state 1100-350}), we notice that this negativeness of the pressure of the $\Omega(1)$ and $\Omega(2)$ substrata is consistent with the observational data of the negativeness of the pressure of dark energy (\cite{ZylaPA2020}, p.\ 490) and with theoretical prediction of the negativeness of the pressure of dark energy \cite{ZeldovichYB1968,SahniV2008}.

According to the theory of vacuum mechanics \cite{WangXS200804}, the $\Omega(1)$ substratum is the origin of the electromagnetic phenomena. Since light is a special kind of electromagnetic wave, the light phenomena also stem from the $\Omega(1)$ substratum. Thus, the existence of the $\Omega(1)$ substratum is supported by electromagnetic phenomena \cite{WangXS200804}. Presently we cannot determine the density $\rho_{1}$ of the $\Omega(1)$ substratum. Therefore, we cannot exclude the possibility that the concept of dark energy is a combination of the $\Omega(1)$ and $\Omega(2)$ substrata.

\section{An opinion on the cosmological constant problem \label{sec 1300}}
The cosmological constant problem arose in the late 1960s \cite{ORaifeartaighC2018}. The demonstration of the Casimir effect \cite{CasimirH1948,SparnaayMJ1957} in the late 1950s had convinced many physicists of the reality of the zero-point energy of the vacuum. In 1968, Y. B. Zeldovich proposed that the scalar field associated with the quantum zero-point energy of the vacuum takes the form of an effective cosmological constant \cite{ZeldovichYB1968,SahniV2008}. Applying the theory of quantum field,    he then suggested a lower bound of the cosmological constant $\Lambda=10^{-6}\mathrm{m}^{-2}$, corresponding to a mass density of $\rho_{\Lambda}=10^{20} \mathrm{kg}\cdot\mathrm{m}^{-3}$ \cite{ZeldovichYB1968,SahniV2008,ORaifeartaighC2018}. However, an observational data of the cosmological constant is $\Lambda=1.088(30)\times 10^{-52}\mathrm{m}^{-2}$ (\cite{ZylaPA2020}, p.\ 138), corresponding to a mass density of $\rho_{\Lambda}=5.831(02)\times 10^{-27} \mathrm{kg}\cdot\mathrm{m}^{-3}$. Therefore, quantum field theory predicted a value of the cosmological constant $\Lambda$ that was $46$ orders of magnitude larger than that observed. This theoretical puzzle is known as the cosmological constant problem \cite{ORaifeartaighC2018}. Furthermore, we are also facing two questions: (1) why the cosmological constant is extremely small; (2) why the cosmological constant has the specific nonzero value \cite{ORaifeartaighC2018}. These two questions are sometimes referred to as the new cosmological constant problem \cite{ORaifeartaighC2018}. A related puzzle is the coincidence problem, i.e. why that the energy contribution of the cosmological constant $\Lambda$ is of the same order of magnitude as that of matter in today's universe \cite{ORaifeartaighC2018}.

By the 1980s, some people thought that an effective cosmological constant can be written as $\Lambda_{\mathrm{eff}}=\Lambda_{0}+\Lambda_{\mathrm{zp}}+\Lambda_{\mathrm{ew}}+\Lambda_{\mathrm{qcd}}$, where $\Lambda_{0}$ is a nonquantum term, $\Lambda_{\mathrm{zp}}$ is a contribution from the zero-point energy of the vacuum, $\Lambda_{\mathrm{ew}}$ is a contribution from the electro-weak phase transition, $\Lambda_{\mathrm{qcd}}$ is a contribution from the quantum chromodynamic phase transition \cite{ORaifeartaighC2018}. Thus, the cosmological constant problem had got worse.

Various solutions have been proposed to solve the cosmological constant problem \cite{BorzouA2018}. The first class of solutions is to modify the theory of gravitation. The second class of solutions is to revise the standard model of particle physics. However, the old and new cosmological constant problems are still open.

In my opinion, the origin of the cosmological constant problem may be that the general theory of relativity is a phenomenological theory of gravity. In the general theory of relativity, the Einstein's equations are assumptions \cite{MollerC1972,WeinbergS1972,MisnerC1973}. Although A. Einstein introduced his new concept of gravitational aether (\cite{Kostro2000}, p.\ 63-113), he did not derive his equations theoretically based on his new concept of the gravitational aether. Although the general theory of relativity is a field theory, the definitions of gravitational fields are not based on continuum mechanics \cite{MollerC1972,WeinbergS1972,MisnerC1973,Eringen1980}. Because of the absence of a medium which will transmit gravitational interactions, the general theory of relativity does not reveal the mechanism of gravity. Thus, the general theory of relativity may be regarded as a phenomenological theory of gravity. Based on this phenomenological theory of gravity, some people thought that all kinds of energy-momentum tensors should appear as source terms in the Einstein's equations (\ref{Einstein 100-200}). It may be this unwarranted belief that leads to the so-called cosmological constant problem.

The gravitational interaction seems to differ in character from other interactions. Except the cosmological constant problem, the existing theories of gravity still face other difficulties. For instance, attempts to reconcile the general theory of relativity and quantum mechanics have met some mathematical difficulties (\cite{MaddoxJ1998}, p.\ 101). Thus, it seems that new ideas about the gravitational phenomena are needed. Following Einstein \cite{EinsteinA2000}, it may be better for us to keep an open and critical mind to explore all possible theories of gravity.

In the history of researches of gravity, there exist some approaches (\cite{ThirringW1961}; \cite{FeynmanRP1995}, page vii; \cite{MisnerC1973}, p.\ 424), which regards Einstein's general theory of relativity as a special relativistic field theory in an unobservable flat spacetime, to derive the Einstein's equations. Recently, C. Overstreet et al. split a cloud of cold rubidium atoms into two atomic wave packets about 25 centimeters apart and subjected one of the wave packets to gravitational interaction with a large mass \cite{OverstreetC2022}. The results show that the potential of gravitational field creates Aharonov-Bohm phase shifts analogous to those produced by the potential of electromagnetic field. Therefore, not only the gravitational fields are real, but also the potentials of gravitational field have real physical influences on the quantum states of matter systems. However, neither these special relativistic field theories (\cite{ThirringW1961}; \cite{FeynmanRP1995}, page vii; \cite{MisnerC1973}, p.\ 424) of gravity nor the general theory of relativity can provide a physical definition of the tensorial potential of gravitational fields.

Inspired by these special relativistic field theories \cite{ThirringW1961} of gravitation, we construct a special relativistic field theory \cite{WangXS201908,WangXS202104}  of gravitation in the Minkowski spacetime based on the theory of vacuum mechanics. Generalized Einstein's equations in some special non-inertial reference frames are derived \cite{WangXS201908,WangXS202104}. If the field is weak and the reference frame is quasi-inertial, these generalized Einstein's equations reduce to Einstein's equations \cite{WangXS201908,WangXS202104}. Thus, this new theory \cite{WangXS201908,WangXS202104} of gravitation may also explain all the experiments which support the general theory of relativity. There exist some differences between this new theory of gravitation \cite{WangXS200810,WangXS201908,WangXS202104} and the general theory of relativity. For instance, in this new theory, gravity is transmitted by the $\Omega (0)$ substratum. The tensorial potential of gravitational field is defined based on special relativistic continuum mechanics \cite{WangXS201908,WangXS202104}. Thus, this new theory \cite{WangXS200810,WangXS201908,WangXS202104} of gravitation is established on the solid bases of the principles of continuum mechanics. According to this new theory of gravitation, the gravitational field is defined by the velocity field of the $\Omega(0)$ substratum \cite{WangXS200810}. Thus, the superposition principle of gravitational fields is a corollary of the superposition theorem of the velocity field of ideal fluids \cite{WangXS200810}. The gravitational mass is defined by a quantity which linearly depends on the strength of a sink flow in the $\Omega(0)$ substratum \cite{WangXS200810}. The tensorial potential of gravitational field is defined by a combination of the tensor of strain rate, the scalar potential and vector potential of the velocity field of the $\Omega(0)$ substratum \cite{WangXS201908}.  The generalized Einstein's equations are derived from the Euler-Lagrange equations of the Lagrangian constructed based on the tensorial potential of gravitational field \cite{WangXS201908,WangXS202104}. The gravitational interactions between two particles are defined by the inverse-square attractive interactions between two sinks caused by fluidic pressure in the $\Omega(0)$ substratum \cite{WangXS200810}. Therefore, the mechanism of gravity may be explained based on the principles of continuum mechanics.

According to the theory of vacuum mechanics \cite{WangXS200810}, two sinks in the $\Omega(0)$ substratum will gravitate with each other. Only those energy-momentum tensors of discrete or continuously distributed sinks in the $\Omega(0)$ substratum are permitted to act as the source terms in the generalized Einstein's equations (\ref{Einstein 150-200}) \cite{WangXS201908,WangXS202104}.

The particles that constitute the $\Omega(1)$ and $\Omega(2)$ substrata are sinks in the $\Omega(0)$ substratum. Thus, the particles of the $\Omega(1)$ and $\Omega(2)$ substrata will also generate gravity \cite{WangXS200810}. Therefore, only three kinds of energy-momentum tensor, i.e. $T^{\mathrm{m}}_{\mu\nu}$, $T^{\Omega(1)}_{\mu\nu}$ and $T^{\Omega(2)}_{\mu\nu}$, are qualified for the source terms in the generalized Einstein's equations (\ref{Einstein 150-200}). Not all kinds of energy-momentum tensors are allowed to act as source terms in the generalized Einstein's equations (\ref{Einstein 150-200}). Therefore, the zero-point energy of electromagnetic fields, the energy from the electro-weak phase transition, the energy from the quantum chromodynamic phase transition, etc., should not act as source terms in the generalized Einstein's equations (\ref{Einstein 150-200}). Thus, the cosmological term $\Lambda g_{\mu\nu}$ may not result from the zero-point energy of electromagnetic fields or other energies.

From Eqs.\  (\ref{Einstein 1100-500}) and Eqs.\  (\ref{relationship 1100-400}), we notice that the origin of the cosmological term $\Lambda g_{\mu\nu}$ in Eqs.\ (\ref{Einstein 100-200}) may be the energy-momentum tensors $T^{\Omega(1)}_{\mu\nu}+T^{\Omega(2)}_{\mu\nu}$ of the $\Omega(1)$ and $\Omega(2)$ substrata. Therefore, the origin of the cosmological constant $\Lambda$ may be the energy-momentum tensors of the $\Omega(1)$ and $\Omega(2)$ substrata. Thus, the smallness of the cosmological constant $\Lambda$ may be explained by the smallness of the energy-momentum tensors of the $\Omega(1)$ and $\Omega(2)$ substrata. Therefore, the old and new cosmological constant problems \cite{ORaifeartaighC2018} may be  explained based on the new theory \cite{WangXS200810,WangXS201908,WangXS202104} of gravitation.

\section{Conclusion \label{sec 1500}}
Presently, we have no methods to determine the density of the $\Omega(1)$ substratum. Therefore, we also suppose that vacuum is filled with a third kind of continuously distributed substance, which may be called the $\Omega(2)$ substratum. Based on a theorem of V. Fock on the mass tensor of a fluid, the contravariant energy-momentum tensors of the $\Omega(1)$ and $\Omega(2)$ substrata are established. Quasi-static solutions of the gravitational field equations in vacuum are obtained. Based on an assumption, relationships between the contravariant energy-momentum tensors of the $\Omega(1)$ and $\Omega(2)$ substrata and the contravariant metric tensor are obtained. Thus, the cosmological constant is calculated  theoretically. The negativeness of the pressure of the $\Omega(1)$ and $\Omega(2)$ substrata is consistent with the negativeness of the pressure of dark energy. Therefore, the $\Omega(1)$ and $\Omega(2)$ substrata may be a possible candidate of the dark energy. According to the theory of vacuum mechanics, there exists an inverse-square attractive force between two sinks in the $\Omega(0)$ substratum. This attractive force may be regarded as the gravitational interaction between elementary particles. Therefore, only those energy-momentum tensors of discrete or continuously distributed sinks in the $\Omega(0)$ substratum are permitted to act as the source terms in the generalized Einstein's equations. Thus, the zero-point energy of electromagnetic fields is not qualified for a source term in the generalized Einstein's equations. In my opinion, the origin of the cosmological constant problem may be that the general theory of relativity is a phenomenological theory of gravity. Some people believed that all kinds of energies should act as source terms in the Einstein's equations. It may be that this unwarranted belief  leads to the cosmological constant problem.

\section*{Appendix: An introduction of the theory of vacuum mechanics}\label{sec 2000}
According to E. Whittaker, Descartes was the first to bring the concept of aether into science by suggesting that it has mechanical properties (\cite{WhittakerE1910}, p.\ 2). Sir I. Newton pointed out that his inverse-square law of gravitation did not touch on the mechanism of gravitation (\cite{WhittakerE1951}, p.\ 28; \cite{HirosigeT1968}, p.\ 91; \cite{CohenIB1980}, p.\ 117). He suggested an explanation of gravity based on the action of an aethereal medium pervading the space (\cite{Newton1718}, p.\ 325). In 1861, in order to obtain a mechanical interpretation of electromagnetic phenomena, Maxwell established a mechanical model of a magneto-electric medium (\cite{SchaffnerKF1972}, p.\ 79-84).

In the years 1905-1916, Einstein abandoned the concept of electromagnetic aether in his theory of relativity (\cite{Kostro2000}, p.\ 27-61). H. A. Lorentz believed that general relativity could be reconciled with the concept of an ether at rest. Einstein changed his view later and introduced his new concept of ether (\cite{Kostro2000}, p.\ 63-113).

Following these researchers, we propose the following mechanical model of the universe \cite{WangXS200804,WangXS200810,WangXS2014PhysicsEssays,WangXS201908,WangXS202104}.

Matter is composed of molecules. Molecules are constructed by atoms. Atoms are formed by elementary particles. Therefore, the universe is composed of elementary particles and vacuum. According to Descartes (\cite{WhittakerE1910}, p.\ 2), every physical phenomenon could be interpreted in the framework of a mechanical model of the universe. Following Descartes' scientific research program, we need to establish a mechanical model of vacuum and another mechanical model of elementary particles.
Thus, we introduce the following assumptions \cite{WangXS200810}.
\begin{assumption_my}\label{assumption 2000-100}
Suppose that vacuum is filled by an extremely thin medium which may be called the $\Omega(0)$ substratum.
\end{assumption_my}

The idea that microscopic particles are sink flows in a fluidic substratum has been proposed by J. C. Maxwell (\cite{WhittakerE1951}, p.\ 243), B. Riemann (\cite{RiemannB2004}, p.\ 507), H. Poincar$\acute{e}$ (\cite{PoincareH1997}, p.\ 171) and J. C. Taylor (\cite{Taylor2001}, p.\ 431-436). Following these researchers, we introduce the following assumption \cite{WangXS200810}.
\begin{assumption_my}\label{assumption 2000-200}
All the elementary particles were made up of a kind of elementary sinks of the $\Omega(0)$ substratum. These elementary sinks were created simultaneously. The initial masses and the strengths of the elementary sinks are the same.
\end{assumption_my}

These basic sinks of the $\Omega(0)$ substratum may be called monads after Leibniz. Based on Assumptions \ref{assumption 2000-100}-\ref{assumption 2000-200} and other auxiliary assumptions, the force $\mathbf{F}_{12}$  exerted on the particle with mass $m_2$ by the velocity field of the $\Omega(0)$ substratum induced by the particle with mass $m_{1}$ is \cite{WangXS200810}
\begin{equation}\label{inverse-square 2000-300}
\mathbf{F}_{12}(t)=- \gamma_{\mathrm{N}}(t)\,\frac{m_{1}(t) m_2(t)}{r^2}\,
\hat{\mathbf{r}}_{12}\,,
\end{equation}
where $\gamma_{\mathrm{N}}(t)=\rho_{0}  q^{2}_{0}/(4\pi m^{2}_{0}(t))$,
$m_{0}(t)$ is the mass of monad at time $t$, $-q_{0}( q_{0} > 0)$ is
the strength of a monad, $m_{1}(t), m_{2}(t)$ are the masses of the two particle at time
$t$, $\rho_{0}$ is the density of the $\Omega(0)$ substratum.

Since the gravitational interactions between sinks in the $\Omega(0)$ substratum are transmitted by this extremely thin fluid, the $\Omega(0)$ substratum may also be called the gravitational aether. As a byproduct, there exists
a universal damping force \cite{WangXS200810}
\vspace*{-2pt}
\begin{equation}\label{damping 2000-500}
\mathbf{F}_{\mathrm{d}} = -\frac{\rho_{0}
q_{0}}{m_{0}}\, m \mathbf{v}
\end{equation}
exerted on each particle, $m$ is the mass of the particle,
$\mathbf{v}$ is the velocity of the particle relative to the
 $\Omega(0)$ substratum.

Thomson's analogies between electrical phenomena and elasticity helped J. C. Maxwell to establish a mechanical model of electrical phenomena (\cite{WhittakerE1951}, p.\ 246). Following J. C. Maxwell, we introduce the following assumption \cite{WangXS200804}.
\begin{assumption_my}\label{assumption 2000-600}
Suppose that vacuum is filled with another kind of substance which may be called the $\Omega(1)$ substratum.
\end{assumption_my}

The particles that constitute the $\Omega(1)$ substratum may be called the $\Omega(1)$ particles. Following Lord Kelvin, we introduce the following assumption \cite{WangXS202104}.
\begin{assumption_my}\label{assumption 2000-700}
The $\Omega(1)$ particles and elementary particles are formed of monads.
\end{assumption_my}

Inspired by J. C. Maxwell (\cite{WhittakerE1951}, p.\ 243), we introduce the following assumption \cite{WangXS200804}.
\begin{assumption_my}\label{assumption 2000-800}
Electric charges in the universe are sources or sinks in the $\Omega(1)$ substratum.
\end{assumption_my}

We may define a source in the $\Omega(1)$ substratum as a negative electric charge and define a sink in the $\Omega(1)$ substratum as a positive electric charge. Based on Assumption \ref{assumption 2000-600}, Assumption \ref{assumption 2000-800} and other auxiliary assumptions, the equation of momentum conservation of the $\Omega(1)$ substratum can be written as \cite{WangXS200804}
\begin{equation}\label{momentum 2000-900}
 W \nabla^2 \mathbf{u}
 + (W +\lambda )\nabla (\nabla \cdot \mathbf{u})
  = \rho_{1} \,\frac{\partial^2 \mathbf{u}}{\partial t^2}
 - \frac{\rho_{\mathrm{e}}\mathbf{v}_{\mathrm{e}}}{k_{\mathrm{Q}}}\,,
\end{equation}
where $W$ is the shear modulus of the $\Omega(1)$ substratum, $\lambda$ is the Lam\'{e} constant, $\mathbf{u}$ is the
displacement vector, $\rho_{1}$ is the density of the $\Omega(1)$ substratum, $\rho_e$ is the density
of electric charges, $\mathbf{v}_{\mathrm{e}}$ is the average speed of electric charges, $k_{\mathrm{Q}}$ is a dimensionless constant, $t$ is time, $\nabla =
\mathbf{i}\partial /\partial x + \mathbf{j}\partial /\partial y + \mathbf{k}\partial /\partial z$ is the Hamilton operator, $\nabla^2$ is the Laplace operator.

It is known that Maxwell's equations in vacuum can be transformed to the following equations (\cite{Jackson1999}, p.\ 240)
\vspace*{-4pt}
\begin{eqnarray}
&&\rule{-1cm}{0pt}\nabla^2 \varphi + \frac{\partial
}{\partial t} (\nabla \cdot \mathbf{A}) = - \frac{\rho_{\mathrm{e}}}{\epsilon_{0}}\,,\label{potential 2000-1010}\\[2pt]
&&\rule{-1cm}{0pt}\nabla^2 \mathbf{A} - \nabla(\nabla \cdot
 \mathbf{A}) - \mu_{0}\epsilon_{0}\frac{\partial
}{\partial t}\left ( \nabla \varphi - \frac{\partial
\mathbf{A}}{\partial t}\right )= -
\mathbf{j}_{\mathrm{e}}\,,\rule[-11pt]{0pt}{0pt}\label{potential 2000-1020}
\end{eqnarray}
where $\varphi$ is the scalar electromagnetic potential, $\mathbf{A}$ is the vector electromagnetic potential, $\rho_{\mathrm{e}}$ is the density field of electric charges, $\mathbf{j}_{\mathrm{e}}$ is the electric current density, $\epsilon_{0}$
is the dielectric constant of vacuum, $\mu_{0}$ is magnetic permeability of vacuum.

The surprising similarity between Eq.\,(\ref{momentum 2000-900}) and Eq.\,(\ref{potential 2000-1020}) can be noticed. Indeed, applying Stokes-Helmholtz resolution theorem of vector field and introducing some auxiliary assumptions, the following Maxwell's equations in vacuum can be derived \cite{WangXS200804}
\begin{eqnarray}
&& \nabla \cdot \mathbf{E}\!\!=\!\!\frac{\rho_{\mathrm{e}}}{\epsilon_{0}}\,, \quad
\nabla \times\mathbf{E} \!\!=\!\! -\frac{\partial \mathbf{B}}{\partial t}\,, \label{Maxwell 2000-1120} \\
&& \nabla \cdot \mathbf{B}\!\!=\!\! 0\,, \quad
\frac{1}{\mu_{0}}\,\nabla \times\mathbf{B}
 \!\!=\!\! \mathbf{j}_{\mathrm{e}}+\epsilon_{0}\,\frac{\partial \mathbf{E}}{\partial t} \,, \label{Maxwell 2000-1140}
\end{eqnarray}
where $\mathbf{E}$ is the electric field vector, $\mathbf{B}$ is the magnetic induction vector.

Since the electromagnetic interactions between sources or sinks in the $\Omega(1)$ substratum are transmitted by this medium, the $\Omega(1)$ substratum may also be called the electromagnetic aether. The electromagnetic aether behaves as a visco-elastic continuum \cite{WangXS200804}. Maxwell's equations approximately describe the macroscopic behaviors of the $\Omega(1)$ particles, in analogy to the way that classical elastic mechanics approximately describes the macroscopic behaviors of the atoms of solid materials.

Eq.\  (\ref{damping 2000-500}) shows that there exists a universal drag force exerted on each sink of the $\Omega(0)$ substratum. Therefore, each monad, each $\Omega(1)$ particle and each microscopic particle, as sinks in the $\Omega(0)$ substratum, will experience a universal drag force. On the other hand, all the monads, $\Omega(1)$ particles and microscopic  particles are experiencing a kind of stochastic force exerted by the $\Omega(0)$ substratum \cite{WangXS2014PhysicsEssays}. Thus, these particles are undertaking some kinds of stochastic movements \cite{WangXS2014PhysicsEssays}. Based on this universal damping force and some auxiliary assumptions, microscopic particles are found to obey a generalized nonrelativistic Schr\"{o}dinger equation \cite{WangXS2014PhysicsEssays}.

Therefore, the gravitational phenomena, the electromagnetic phenomena and the nonrelativistic quantum phenomena are explained self-consistently in a unified theory. For convenience, we may call these theories \cite{WangXS200804,WangXS200810,WangXS2014PhysicsEssays,WangXS201908,WangXS202104} as the theory of vacuum mechanics.

It is interesting to notice that the gravitational phenomena and the nonrelativistic quantum phenomena both stem from the interactions between particles and the $\Omega(0)$ substratum. Gravity and stochasticity are two intrinsic characters of all microscopic particles.

There exist some unsolved theoretical and experimental problems in the fields of vacuum mechanics and other related fields. For instance, what is the physical meaning of the sources or sinks of the $\Omega(2)$ substratum? Is it possible for us to detect some of the predictions of the theory of vacuum mechanics by experiments or observations?

\end{document}